\documentstyle[epsf]{elsart}

\def\beq         {\begin{equation} }
\def\eeq         {\end{equation} }
\def\Fst         {Fock--space trun\-ca\-tion}
\def\cs          {colour--singlet}
\def\lc          {light--cone}
\def\Fs          {Fock--space}
\def\ost         {one--sector trun\-ca\-tion}
\def\tst         {two--sector trun\-ca\-tion}
\def\len         {light--cone ener\-gy}
\def\lmo         {light--cone mo\-men\-tum}
\def\gs          {ground state}
\def\ms          {mass spectrum}
\def\Col         {Coulomb}
\def\DLCQ        {Dis\-cre\-tised Light--Cone Quan\-ti\-sa\-tion}
\def\LFTD        {Light--Front Tamm--Dan\-coff}
\def\fof         {front form}
\def\TD          {Tamm--Dan\-coff}
\def\Mts         {\widetilde{M}^2}
\begin{document}

\pagenumbering{roman}
\begin{frontmatter}
\title{
{\bf {Constituent Quark Picture out of QCD in two--dimensions ---
on the Light--Cone}}}
\author{M. Heyssler\thanksref{email1}} and
\author{A.C. Kalloniatis}
\address{ \em
Max--Planck--Institut f\"ur Kernphysik\\
Postfach 10 39 80\\
D--69029 Heidelberg, Germany}
\thanks[email1]{from May 1, 1995:\\
University of Durham, Department of Physics, Durham DH1 3LE, U.K. }

{\small Preprint: {\bf MPIH--V25--1994} (Revised)}

\begin{abstract}
Using DLCQ as a nonperturbative method, we test
{\Fs} truncations in ${\rm QCD}_{1+1}$
by studying the mass spectra of hadrons in colour
SU(2) and SU(3) at finite harmonic resolution $K$.
We include $q\bar q q\bar q$ states for mesons
and up to $qqq q\bar q$ states for baryons.
With this truncation, we
give `predictions' for the masses of the first five states
where finite $K$ effects are minimal.
\end{abstract}

\end{frontmatter}

\newpage

\pagenumbering{arabic}
\begin{frontmatter}

\title{
{\bf {Constituent Quark Picture out of QCD in two--dimensions ---
on the Light--Cone}}}
\author{M. Heyssler\thanksref{email2}} and
\author{A.C. Kalloniatis}
\address{ \em
Max--Planck--Institut f\"ur Kernphysik\\
Postfach 10 39 80\\
D--69029 Heidelberg, Germany}
\thanks[email2]{from May 1, 1995:\\
University of Durham, Department of Physics, Durham DH1 3LE, U.K. }

\begin{abstract}
Using DLCQ as a nonperturbative method, we test
{\Fs} truncations in ${\rm QCD}_{1+1}$
by studying the mass spectra of hadrons in colour
SU(2) and SU(3) at finite harmonic resolution $K$.
We include $q\bar q q\bar q$ states for mesons
and up to $qqq q\bar q$ states for baryons.
We give `predictions' for the masses of the first five states
where finite $K$ effects are minimal.
\end{abstract}


\end{frontmatter}

\underline{Introduction.}
The idea of formulating mechanics and field theory on null--plane
surfaces was first stimulated by Dirac in 1949 \cite{Dir49}
with his introduction of the three independent
schemes of Hamiltonian dynamics: the (conventional) instant
form, the {\fof} and the point form. The second
of these schemes was exploited in 1985 by Pauli and Brodsky
\cite{Pau85} in their formulation of
{\em\DLCQ} (DLCQ).
DLCQ is a non--perturbative
Hamiltonian field theoretic method.
The hope is that within this precise treatment
of Quantum Chromodynamics (QCD) something like
a constituent quark or parton picture of hadrons can emerge.
The ingredients for this are first that a unique vacuum
state can be found and second that the low energy hadron
spectrum can be described in terms of a low number
of quark (and gluon) excitations above this state. The first
condition is rigorously satisfied in the {\fof}
if zero modes are ignored \cite{Sch91}.
The second follows from the related positive definiteness of
the {\lmo} operator. It is the concrete realisation
of this in a spectrum which we examine here in QCD in one space and one time
dimension (${\rm QCD}_{1+1}$). This theory has the advantage of
being superrenormalisable.

In DLCQ space is a `box'; in 1+1 dimensions
$x^-=\frac{1}{\sqrt{2}}\left( x^0 - x^1 \right)$ is
restricted to a finite interval
of length $2L$.
Choosing appropriate boundary conditions discretises the
Fourier momenta. Thus the field theory problem is reduced
to one of finite dimensional matrices which can be directly
diagonalised \cite{Sch91}.
The first application to ${\rm QCD}_{1+1}$ was by
Hornbostel in \cite{Hor90}. Mass spectra and wavefunctions
of baryons and mesons were numerically obtained for
finite harmonic resolution $K=\frac{L}{\pi}P^+$ where $P^+$ is the total
{\lmo}. $K$ is dimensionless and itself regulates the
size of the {\Fs}.
The continuum limit, $L\rightarrow \infty$, must be achieved
maintaining finite momentum $P^+$ thus
$K$ must become `infinite', namely large enough to enable
reasonable extrapolation. This can make overwhelming demands
on CPU time.
So a second advantage of a {\Fst}
lies in the computer time and memory saved for numerical calculations.

{\Fst}, also called `{\TD} truncation',
has already seen some justification.
Ground state hadrons in ${\rm QCD}_{1+1}$ have been shown in
\cite{Hor90} to consist only of a very low number of particles:
the {\gs} of an SU($N$) meson can be described by a single
quark--antiquark ($q\bar q$) pair and the {\gs} of an SU($N$)
baryon simply by $N$ quarks. This minimal {\Fs} truncation
is already known to breakdown even for the first excited state.
For example, the first excited state of
an SU($N$) meson was recently
calculated analytically in \cite{Sug94} using the {\LFTD}
(LFTD) method of \cite{Wilson}, a related approach exploiting
the advantages of the {\fof} with a {\Fs} truncation.
They worked in `next to leading' order in {\Fs} truncation.

One aim of the present work is to show that, within DLCQ, the
same level of truncation works {\em not only for the first} but for a
number of the lowest lying excited states.
In SU(2) and SU(3) colour group gauge theory
we will find that a truncation
to a low number of particles reproduces
a large number of the lowest lying states
to a good approximation with
calculations taking about one hour CPU time.
There is a so--called `zero mode problem' in the {\fof}.
Here we ignore these modes to study
the physics of the trivial vacuum, reasonable here since
for finite $N$, SU($N$) gauge theory has no symmetry--breaking
in (1+1) dimensions.

\underline{DLCQ in Brief.}
We shall be cursory in our presentation of the
basic elements of DLCQ as the literature is now quite
extensive, for example \cite{Sch91,Hor90,Hey94}.
The {\lc} coordinate convention is for any Lorentz vector $V^{\mu}$ define
$V^{\pm}=V_{\mp}=\frac{1}{\sqrt{2}}\left( V^0 \pm V^1\right)$.
We start with the Lagrangian of QCD with one quark flavour
\beq
{\cal L} = - \frac{1}{4}F^{a\mu\nu}F^a_{\mu\nu} + \frac{1}{2} \left(
          \overline{\Psi}i\gamma^{\mu}D_{\mu}\Psi + {\rm h.c.}\right)
          - m\overline{\Psi}\Psi , \label{lagrangian}
\eeq
with the field--strength tensor $F^a_{\mu\nu}=\partial_{\mu}A^a_{\nu}-
\partial_{\nu}A^a_{\mu} + g f^{abc}A^b_{\mu}A^c_{\nu}$ and the covariant
derivative $D_{\mu} = \partial_{\mu}{\bf 1} - igA^a_{\mu}{\bf T}^a$ in
the fundamental representation.
Thus the colour matrices
${\bf T}^a$ are related to the Pauli matrices for SU(2) and to the
Gell--Mann matrices for SU(3).
For the ${\gamma}$--matrices we choose the chiral representation
\cite{Itz87}.
The quark field $\Psi$ is just a two component spinor in two dimensions
$\Psi = \left( \Psi_{L,c_i},\Psi_{R,c_i} \right)^{\bf t}$,
where $L$ and $R$ represent chirality and $c_i$ colour.
In one space and one time dimension there is no spin.

We solve this theory
by addressing the eigenvalue equation \cite{Pau85}
\beq
2P^+P^-|\Psi\rangle = M^2|\Psi\rangle.
\label{eigenvalue}
\eeq
Here $P^+, P^-$ are the Poincar\'e generators of respectively
space and time translations: the {\lmo} and the {\len}.
The eigenvalue $M^2$ is the Lorentz invariant mass--squared of the
eigenstate $|\Psi\rangle$.
The link between (\ref{lagrangian}) and (\ref{eigenvalue})
is the energy--momentum tensor. Since the generators are constants of
the motion under evolution in {\it{\lc} time} $x^+$
they can be written in terms of the independent
fields specified at a given time, say $x^+ = 0$.
Ignoring zero modes\footnote{
The zero mode
$A^{+a}_0 \equiv\frac{1}{2L} \int_{-L}^{+L} {\d x^-} A^{+a}(x^-)$
leads to the gauge invariant Wilson loop around
{\lc} space $x^-$ \cite{Man85}. So it cannot actually be gauged away.
Its role has been explored elsewhere
e.g. \cite{Kal94,KPP94}.},
we choose the {\lc} gauge $A^{+a}=0$.
Thus there is only one independent field:
the fermion component $\Psi_{R,c_i}$.
Both the gluon field $A^{-a}$ and the
other fermion component are constrained by, respectively,
their Maxwell and Dirac equations which are trivially
implemented.
Quantisation is achieved by imposing
anticommutation relations on this independent field
$\{ \Psi_{R,c_i}(0,x^-),\Psi^{\dagger c_j}_{R}(0,y^-)\}=
\frac{1}{\sqrt{2}}\delta^{c_j}_{c_i}\delta(x^--y^-)$
for a fixed {\lc} time $x^+=y^+=0$.
Introducing the plane wave expansion as `initial data'
\beq
\Psi_{R,c_i}(0,x^-)
=\frac{1}{\sqrt[4]{2}}\frac{1}{\sqrt{2L}}
\sum\limits_{n}^{\Lambda}
\left( b_{n,c_i}e^{-ik_n^+x^-} + d^{\dagger}_{n,c_i}e^{+ik_n^+x^-} \right),
\label{expansion}
\eeq
we obtain as the non--vanishing anticommutators
$\{b^{\dagger c_i}_n, b_{m,c_j} \} = \{ d^{\dagger}_{n,c_j},d_m^{c_i}\}
=\delta^{c_i}_{c_j}\delta_{n,m}$.
As in \cite{Hor90}, we impose antiperiodic
boundary conditions for the fermion field:
$\Psi_{R,c_i}(x^-+2L) = -\Psi_{R,c_i}(x^-)$.
This gives
$n\in\{\frac{1}{2},\frac{3}{2},...,\frac{\Lambda}{2}\}$.
The cut--off
${\Lambda}$ drops out after normal ordering,
reflecting superrenormalisability.
The generators $P^{\pm}$ can thus be
expressed in terms of the Fock--modes $b_{n,c_i}$ and $d_{n,c_i}$.
Then $:P^+:$ is just proportional to the number
operator in quarks and antiquarks even in the interacting theory
and $:P^-:$ has a kinetic term also proportional
to the number operator and an interaction term bilinear in
the quark current $j^{+a}(x^-)=\frac{2}g\Psi^{\dagger c_i}_R
\left( T^a \right)_{c_i}^{c_j}\Psi_{R,c_j}$.
The detailed expression for
the interaction term can be found in \cite{Hor90,Hey94}. Its main
feature is the linear {\Col} potential between
the quark currents obtained from elimination of $A^{-a}$.

We now discuss how the wavefunctions are represented in the Fock--basis.
For a given colour group SU($N$), the {\cs} state
$|\Psi\rangle$ depends on the baryon number $B$ and the
harmonic resolution $K$. Group theoretic aspects of the following
procedure can be found in \cite{Ham62}.
To construct an  SU($N$) meson state ($B=0$) at a fixed
harmonic resolution $K$, we {\em begin} with a two particle
{\cs} state:
$
|{\rm meson}\rangle = \delta^{c_2}_{c_1}b^{\dagger c_1}_{n_1}
                      d^{\dagger}_{n_2,c_2}|0\rangle.
$
To this we can append {\cs} $q\bar q$ creation operators
with equal or increasing momentum $n$ until the total
momentum $K$ is saturated by the sum of the parton
momenta $\sum_i n_i$. Thus the resulting state
must be multiplied by a Kronecker $\delta_{K,\sum_i n_i}$.
For an SU($N$) one--baryon state ($B=1$) we proceed
analogously. As a basic state we contract
$N$ quarks with the antisymmetric epsilon--tensor of
rank $N$:
$
|{\rm baryon}\rangle = \epsilon_{c_1c_2\cdots c_N}
b^{\dagger c_1}_{n_1}b^{\dagger c_2}_{n_2}\cdots
b^{\dagger c_N}_{n_N}|0\rangle.
$
Now, as above, we can append as many $B=0$ operators, $q\bar q$ pairs,
as the total momentum $K$ allows.
The Hilbert space so constructed is overcomplete and not orthonormal.
This is dealt with in the code as in
\cite{Hor90} by weeding out states with zero inner product
$\langle i | j \rangle$.

These are the tools for directly
solving Eq.~(\ref{eigenvalue}).
For this purpose a computer code was set up to construct the {\Fs}
and calculate $\widetilde{M}^2$
directly for given $K$, $N$ and $B$. The dimensionless parameter
$\lambda=\left( 1+\pi m^2/g^2 \right)^{-1/2} \in [0,1]$ allows
features of the spectrum over an entire range of couplings or
masses to be seen in one finite domain plot.
Eventually, Eq.~(\ref{eigenvalue})
can be rewritten purely in terms of dimensionless quantities as
\beq
\Mts|\Psi\rangle \equiv
\frac{ \pi M^2/g^2}{1 + \pi m^2 / g^2} |\Psi\rangle =
(1-\lambda^2)K\widehat{T}|\Psi\rangle + \lambda^2 K \widehat{V}|\Psi\rangle,
\label{eigenvaluenew}
\eeq
where $\Mts$ is a {\em dimensionless} invariant mass--squared,
and $\widehat{T}$ and $\widehat{V}$ are sums over the Fock--modes with all
dimensionful quantities, such as $m, g$ and $L$, stripped off. Their detailed
form can be found in \cite{Hor90,Hey94}.
For historical reasons, the spectrum is normalised
in units of $g/\sqrt{\pi}$ which is
the mass of the lowest boson in the Schwinger model though no such
state exists in the massless fermion limit of this theory.

\underline{{\Fs} Truncation in a Typical DLCQ Spectrum.}
To be concrete, consider the {\ms} with full {\Fs}
allowed at fixed momentum $K=33/2$.
This value was chosen because the lowest lying states in the spectrum
became stable for various couplings in this region of $K$,
but remained computable in a reasonable amount
of CPU time (the reader is referred to \cite{Hey94} for details).
For an SU(3) baryon $K=33/2$ means
storing 812 states and thus
calculating $812^2$ matrix elements.
This procedure, on a DEC OSF/1 V1.3, takes
about 70 minutes the least part being devoted to dia\-go\-na\-li\-sa\-tion.
This time can be further reduced by a factor of
{\it three} by {\Fst}.
The {\Fs} truncations we use in this letter are defined in Table~1.

A {\Fst} which includes only
sector 1 is a {\em \ost}.
Taking sectors 1 and 2 into account is called a {\em \tst}.
When we say a hadron state is well--described by a particular
sector truncation we mean two things:
(1) that the invariant mass--squared in the
truncated approximation, denoted
$\Mts_{\rm tru}$,
agrees with that calculated
with the full available {\Fs} for finite $K$,
and (2) that the
result agrees with other methods, such as semi--analytic \cite{tHo}
or lattice \cite{Ham}, where such
data is available and reliable.
As mentioned \cite{Hor90},
the {\gs} of each SU($N$) hadron is well--described by a {\ost}.
This has been cross--checked in \cite{Hey94}
where it is also clear that {\it excited states} are
{\it poorly} described in this approximation.
This is a consequence of the absence of interactions
between the different {\Fs} sectors. Some of these interactions are
implemented when the second {\Fs} sector is taken into
account.

Let us turn then to a typical excitation spectrum computed with DLCQ
in the {\it two--sector} truncation.
There is no need to show both mesons and baryons
for SU(2) and SU(3) as they are all qualitatively similar.
We just give as an example in Fig.~1 only the SU(3) baryon spectrum.

In such plots there are places where
finite $K$ artifacts are dominant: in particular the mass gaps
between the bunches of states at low $\lambda$ and the mass gap at
$\lambda = 1$. This was seen by studying how the gaps change for
varying $K$. The detailed reasons for these artifacts
we come to later.
But not all the gaps are artifacts: those between the lowest
states for intermediate $\lambda$ values are stable against
variation in $K$. Another reason to regard these gaps as
physical is that they
{\it emerge from the continuum} spectrum which these lowest lying
states already {\it roughly} form in the free theory limit $\lambda = 0$.
Thus this low energy regime is one where
the real physics, well understood
already in \cite{Hor90},
is perceptible even at finite $K$. The physics is dominated by the
linear {\Col} potential giving increasing repulsion
(with increasing $\lambda$ or $g$) between partons in the baryon.
On the other hand, with
a {\tst} pair--production terms are allowed in the interaction
\cite{Hor90,Hey94}. Thus the invariant mass of the
hadron should eventually {\em decrease} as
$\lambda \rightarrow 1$ since the {\Col} energy is lost into
the production of more partons.
This we observe
except for states with $\Mts_{\rm tru}>80$
and coupling $\lambda>0.6$. This is, in
this case, a consequence of the {\Fst}: no further partons
can be created to absorb the increasing {\Col} energy. Thus
we encounter the first place where the {\Fst} has
harmed the physics. Even without
this kind of {\Fst}, this would still occur, maybe at
higher states, for any finite $K$, as $K$ itself is a truncation
in the number of allowed partons.
At $\lambda =1$ we see a group of the lowest states going to zero.
This is the massless fermion limit being manifested: in 1+1
dimensions these particles are in colinear motion, so one can
always find a frame in which the total mass of the system
is zero.
As mentioned, the finite $K$ artifacts are mainly at the extreme regimes in
$\lambda$ -- in the gaps at $\lambda = 0$ and $\lambda = 1$.
With higher $K$ these turn into a continuum indicating the
continuous momentum that can be assigned to
free or infinitely coupled constituent quarks.

As mentioned,
for $0.3<\lambda < 1.0$ we find a mass gap between the {\gs} and the
first excited state in the SU(3) baryon. This seems stable as
$K\rightarrow\infty$
\cite{Hor90,Hey94}. In this region we propose that real physics
is best approximated for finite $K$. However, how do
these results in this region compare with those from a calculation
at `full' (at given $K$) {\Fs}?

\underline{Success of Truncation at Finite $K$.}
We now make a quantitative comparison between the {\tst} and
the full {\Fs} for fixed momentum $K$.
The actual spectra are not presented here but can
be found in \cite{Hey94}. We define the
quantity
$\Delta_i=\Mts_{{\rm tru},i} - \Mts_i$,
which is the difference between the mass--squared
of the $i$th state of the two--sector calculation and the
corresponding quantity from
the full calculation. This difference is shown
in Fig.~2 for the first 20 states over the whole range of $\lambda$
for the SU(2) and SU(3) hadrons.

The most obvious feature in these plots is the `valley' in Fig.~2c
corresponding to the SU(3) meson. This occurs in the
non--interacting theory limit.
The effect is due to the removal of a lowest energy
$q \bar{q} q \bar{q} q \bar{q}$ state as a result of the two--sector
approximation\footnote{
We have done the calculation with a
truncation at {\em six} particles and observed the
valley to vanish with maximum absolute deviation $\Delta_{20} = 0.5$.}.
The energy of this missing state in the free theory is slightly above
some of the $q\bar{q}q\bar{q}$ states. Because $K$ is finite there is
no density of states around it that would better match the energy of these
neighbouring $q\bar{q}q\bar{q}$ states once this state is
removed in the two-sector truncation. In the truncated space spectrum
this missing state is replaced as the seventh state
by a $q\bar{q}q\bar{q}$ state
with slightly lower energy so that the {\it difference}
$\Delta_{7}$ is negative. Hence the negative valley.
Some higher states are of course degenerate in the full-calculation
and remain in degeneracy in the two-sector truncation. Thus they remain
comparable in energy and the valley disappears beyond the seventh state
until a similar thing occurs again at the fifteenth state.
With higher $K$ all the states in the spectrum
would be closer together and the
removal of a higher {\Fs} state would not cause such mismatches.

In general then, we can say at worst the lowest six, and at best the
lowest ten, states agree for the two choices of {\Fs} size
in the whole regime of $\lambda$
by a {\tst}. Even for moderate values
of $\lambda$ ($\lambda<0.6$) the lowest 20 states fit perfectly.
Evidently in the
strong coupling regime the numerics for the higher states in a {\tst}
are poor.
The reason for this behaviour has been discussed above:
suppression of pair--production by the {\Fst}.
For the SU(2) hadrons the maximum
difference for the highest considered state ($\Delta=14$) means
a relative difference compared to the full calculation of less
than 8\% .
For the SU(3) hadrons the maximum difference
($\Delta=22$) means less than 10\% relative difference.

The conclusion of this analysis is that
at least for the first five excited states in a given
baryon number spectrum there is a broad range
of couplings for which the truncated {\Fs}
computation is consistent at least with
the full calculation. We next give some
actual numbers for invariant masses.

\underline{Numerical Results for First 5 Mass Eigenstates.}
The numerical results of our {\tst} are presented in Table~2 for the
lowest five excited states and the {\gs} for moderately large
values of dimensionless coupling which we indicate
in terms of $\lambda$ for two cases: the SU(2) meson
and the SU(3) baryon. The intuitive picture this approximation would seek
to justify is: for the meson of a pair of $q \bar{q}$ states in
relative motion, for the baryon of {\gs} $qqq$ state
combining with a $q\bar{q}$ pair. To what extent is this a valid
physical picture for these states?
To judge we compare our results for the {\gs} with those of \cite{Ham}
for coupling $\lambda =$  0.82, 0.58 and 0.33.
Hamer \cite{Ham} gets\footnote{Hamer \cite{Ham} worked in units of $m/g$
and $M/g$.
We have simply converted his results into our units.}
respectively $\widetilde{M}_0^2$ = 2.46, 3.90 and 4.21. We get
$\widetilde{M}^2_{\rm tru,0}=$ 2.33, 3.86 and 4.19 as shown
in Table~2. The comparison is reasonable.
It must be pointed out that
Hamer \cite{Ham} worked in the continuum limit ($K\rightarrow\infty$ in
our case) by extrapolating his values obtained for a finite
grid size via a Pad{\'e} method.
It may seem inappropriate to compare
Hamer's continuum result \cite{Ham}
to our results at finite $K$. We have checked that for
very small fermion mass $m$, indeed our {\gs}
results are poor compared to the lattice
result\footnote{
Our results are about 50\% smaller
than the continuum limit results of \cite{Ham}.}.
However,
as stated above, for intermediate
values of the coupling,
$0.33<\lambda <0.82$ (corresponding to $0.40<m/g<1.6$), the
comparison is excellent. In other words
finite volume artifacts are small in this
range of coupling for the lowest lying states.
Alternately one could
correctly regard finite $K$ as a truncation
of the {\Fs} {\em a priori}, and thus
the question becomes: how good is that as
an approximation to the continuum limit?
Again, for the given coupling range it is
very good.

For large $\lambda$ or small fermion mass
$m$ the finite volume effects grow, as
is also known from the Schwinger model
\cite{Ell88}. Extrapolation using results
over a range of $K$ can be applied to
improve the DLCQ results, as will be shown in a following
paper \cite{Hey95}. But here quantitative comparison
with the results of {\LFTD}
\cite{Sug94} for
ground states and first excited states
can be made.
Our extrapolation results and
comparison to these authors
will be presented elsewhere \cite{Hey95}.

The remaining values in Table~2 for the
five excited states are presented in order
for future comparison with other methods such
as lattice gauge theory or LFTD. The
inherent advantage of DLCQ is that a
huge portion of the excited state spectrum
is obtained with the same (and relatively
small) computational
effort that goes into the {\gs}.
This we believe is DLCQ's advantage over
other methods. A flaw may be that the
plane wave basis implicit in DLCQ may
be too restricted a class of basis functions
to give the higher states with any
reasonable accuracy.
Nonetheless, the plots of Fig.~2 give
some cause for confidence in the results
of Table~2: there is {\em internal consistency}
in the results of DLCQ. Beyond the fifth state
there is some reason to put less trust in the
accuracy of a {\tst} at such a
low $K=33/2$, as
evidenced by the anomalous valley
discussed earlier.
However, the {\Fst} now permits
calculations at {\it higher} $K$ followed by
an {\it extrapolation} to the continuum within reasonable
CPU time. This is treated in a separate work \cite{Hey95}.

\underline{Conclusions.}
We have used DLCQ to compute the spectrum of invariant
masses of mesons and baryons in one--flavour SU(2) and SU(3) gauge
theory in (1+1) dimensions. With a {\tst}
on the {\Fs} we find excellent agreement for the
masses as compared to a full {\Fs} computation for
the first 20 states.
For a region of coupling ranging
from moderately weak to strong we find an absence of
finite $K$ artifacts enabling us to make numerical
`predictions' for up to the first six states in the
SU(2) meson and SU(3) baryon spectra. Hopefully
comparable computations from lattice or LFTD will
be done in the near future enabling a check of these
numbers.
For QCD in 3+1 dimensions one expects chiral symmetry--breaking
to impact on the spectrum of the
lighter mesons. The neglected zero modes may play a pivotal role in this.
More seriously, renormalisation is a significant problem to
be overcome in higher dimensions. Finally,
confinement--generation in 3+1 dimensions remains a big unknown in a
DLCQ treatment.
Modulo these features,
it is a reasonable question to ask how much of the above
insight should work, say, for the intermediate physical mesons
where the constituent quark model is known to work well.
The key mechanism in our spectra was played by the
linearly confining potential. If such a potential were accessible in
3+1 dimensions via DLCQ treatment of QCD, then there is
hope that {\Fs} truncation should work. In other words,
a similarly intuitive picture of the intermediate mass
hadrons consistent with the parton model would emerge but
expressed in terms of fundamentally {\it QCD degrees of freedom}.
The method deserves further
examination in more complicated theories.

\ack
{
The authors would like to thank the following for critical
discussions and suggestions: S.~G\"ullenstern,
L.~Hollenberg, K.~Hornbostel and J.~Vary. B.~van de Sande
is thanked for a critical reading of this manuscript. We
especially thank H.--C.~Pauli for suggesting and supporting the work
throughout.
(ACK) was supported by the DFG under grant PaZ/1--2,
and an MPG Stipendium.
}



\newpage

\begin{table}[t]
\centering
\caption
{
{\em Definition of the two distinct {\Fs} sectors under consideration
for SU($2$) and SU($3$) mesons ($B=0$) and baryons ($B=1$).}
}

\vspace{1.0cm}

\begin{tabular}{|c||c|c||c|c|}\hline
& \multicolumn{2}{|c||}{\rule[-3mm]{0mm}{8mm}\bf SU(2)}
& \multicolumn{2}{|c|}{\rule[-3mm]{0mm}{8mm}\bf SU(3)} \\
\cline{2-5}
{\rule[-3mm]{0mm}{8mm}{\rm sector}} & $B=0$ & $B=1$ & $B=0$ & $B=1$ \\
\hline\hline
{\rule[-3mm]{0mm}{8mm}$1$} &
$q\bar q$ & $qq$ & $q\bar q$ & $qqq$ \\ \hline
{\rule[-3mm]{0mm}{8mm}$2$} &
$q\bar q q\bar q$ & $qq\, q\bar q$ & $q\bar q q\bar q$ & $qqq\,
q\bar q$
\\ \hline
\end{tabular}
\end{table}

\newpage

\begin{table}[t]
\centering

\caption{
Numerical results of our {\tst} calculation for the {\gs} and
the lowest five excited states. The results are shown
for the SU($2$) meson and the SU($3$) baryon.
All entries are in units of $\Mts_{{\rm tru},i}$ as defined by
Eq.~(4).
The values of coupling are presented in
dimensionless units of $\lambda$.}

\vspace{1.0cm}

\begin{tabular}{|c||r@{.}l|r@{.}l|r@{.}l|r@{.}l|r@{.}l|r@{.}l|}\hline
\multicolumn{13}{|c|}{\rule[-3mm]{0mm}{8mm}\bf SU(2) meson}\\ \hline
$\quad\lambda\quad$ &
{\rule[-3mm]{0mm}{8mm} 0}&82 & 0&58 & 0&33 & 0&28 & 0&18 & 0&14 \\ \hline\hline
$\Mts_{{\rm tru},0}$ & 2&33 &  3&86 & 4&19 & 4&26 & 4&21 & 4&17 \\ \hline
$\Mts_{{\rm tru},1}$ & 5&14 &  6&40 & 4&46 & 5&21 & 4&69 & 4&48 \\ \hline
$\Mts_{{\rm tru},2}$ & 7&47 &  8&40 & 6&18 & 5&79 & 5&00 & 4&75 \\ \hline
$\Mts_{{\rm tru},3}$ & 8&53 & 10&09 & 6&77 & 6&18 & 5&07 & 4&79 \\ \hline
$\Mts_{{\rm tru},4}$ & 8&79 & 11&44 & 7&30 & 6&73 & 5&86 & 5&65 \\ \hline
$\Mts_{{\rm tru},5}$ & 9&05 & 12&43 & 7&37 & 6&76 & 5&88 & 5&66 \\ \hline
\multicolumn{13}{|c|}{\rule[-3mm]{0mm}{8mm}\bf SU(3) baryon}\\ \hline
$\quad\lambda\quad$ &
{\rule[-3mm]{0mm}{8mm} 0}&82 & 0&58 & 0&33 & 0&28 & 0&18 & 0&14 \\ \hline\hline
$\Mts_{{\rm tru},0}$ &  5&41 &  9&83 & 10&24 & 10&25 &  9&86 &  9&65 \\ \hline
$\Mts_{{\rm tru},1}$ & 12&78 & 17&25 & 13&62 & 12&84 & 10&95 & 10&31 \\ \hline
$\Mts_{{\rm tru},2}$ & 15&01 & 21&50 & 15&08 & 13&89 & 11&76 & 11&17 \\ \hline
$\Mts_{{\rm tru},3}$ & 15&17 & 22&46 & 15&40 & 14&41 & 12&79 & 12&36 \\ \hline
$\Mts_{{\rm tru},4}$ & 15&63 & 24&76 & 16&47 & 15&49 & 13&39 & 12&77 \\ \hline
$\Mts_{{\rm tru},5}$ & 15&97 & 26&00 & 17&29 & 15&84 & 13&89 & 13&45 \\ \hline
\end{tabular}
\end{table}

\newpage

\begin{figure}[t]
\unitlength1cm
\centering
\begin{picture}(12,9)
\makebox(12,9)[t]{\epsfxsize=12cm
               \epsfysize=9cm
               \epsffile{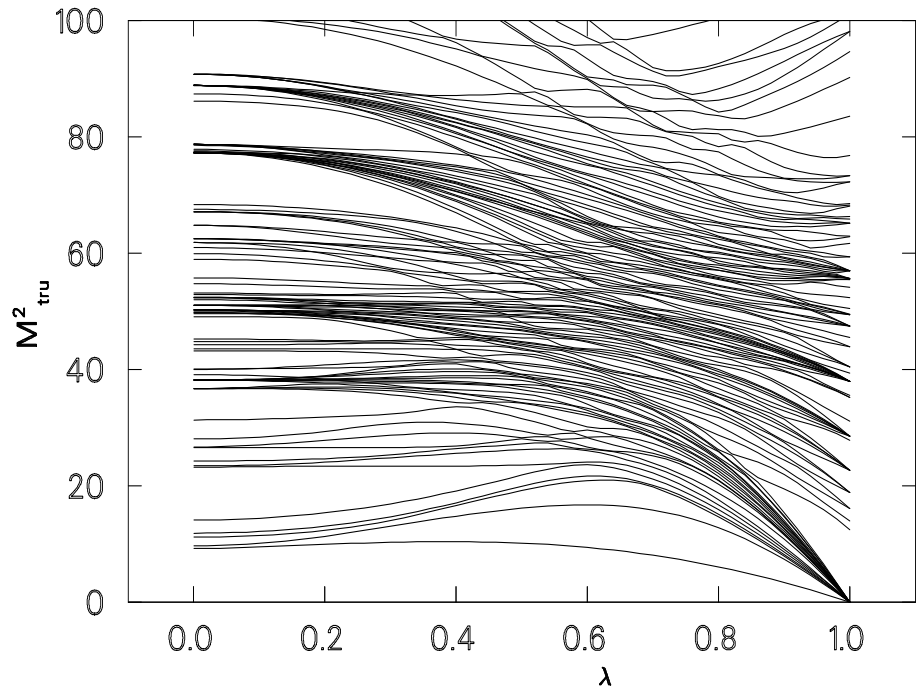}}
\end{picture}

\vspace{1.0cm}

\caption
{
{\bf SU(3) baryon mass spectrum in the two--sector
trunca\-tion.}
\em The {\ms} of the SU($3$) baryon with a
{\tst} defined in Table~1. The units are the
dimensionless mass $\Mts_{\rm tru}$ versus
the dimensionless coupling $\lambda$. The harmonic
resolution used was $K=33/2$.
}
\end{figure}

\newpage

\begin{figure}[t]
\unitlength1cm
\begin{picture}(15,12)
\centering
\makebox(15,12)[t]{\epsfxsize=15cm
                   \epsfysize=12cm
                   \epsffile{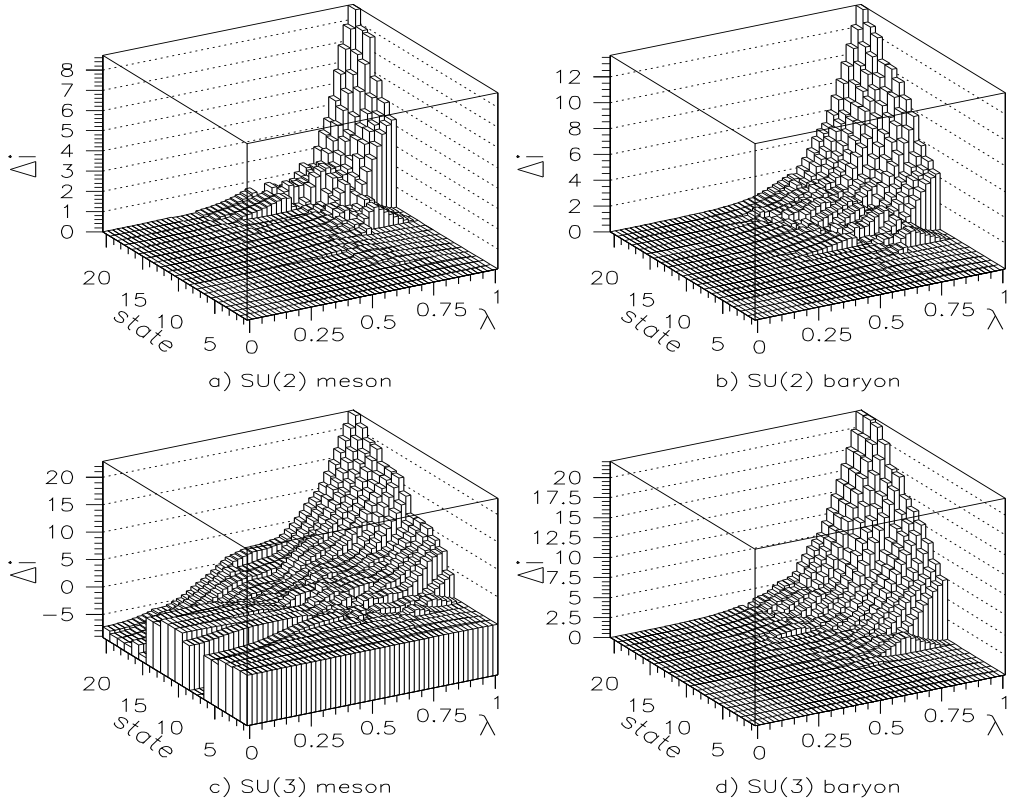}}
\end{picture}

\vspace{0.5cm}

\caption
{
{\bf Difference of the mass spectra in a {\tst}
and the full calculation. }
\em The dimensionless difference $\Delta_i$
for the first $20$ states is plotted over all
$\lambda$. The harmonic resolution
for both the {\tst} and the full {\Fs} is in the cases a)--c) $K=16$,
for the SU($3$) baryon in d) $K=33/2$.
}
\end{figure}

\end{document}